\documentclass[]{spie}  

\usepackage{amsmath,amsfonts,amssymb}
\usepackage{graphicx}
\usepackage[colorlinks=true, allcolors=blue]{hyperref}
\usepackage{lineno}
\usepackage{hyphenat}
\usepackage{subcaption}

\usepackage{hyperref}

\title{The Anti-Coincidence Detector Subsystem for ComPair}

\author[a,b,e]{Zachary Metzler}
\author[c,b,e]{Nicholas Cannady}
\author[d]{Daniel Shy}
\author[b]{Regina Caputo}
\author[b]{Carolyn Kierans}
\author[d]{Richard Woolf}
\affil[a]{University of Maryland, College Park, College Park, MD 20742}
\affil[b]{NASA Goddard Space Flight Center, Greenbelt, MD 20771}
\affil[c]{University of Maryland, Baltimore County, Baltimore, MD 21250}
\affil[d]{United States Naval Research Laboratory, Washington, DC 20375}
\affil[e]{Center for Research and Exploration in Space Science and Technology, NASA/GSFC, Greenbelt, MD 20771}

\authorinfo{Further author information: (Send correspondence to Z.M.)\\Z.M.: E-mail: zmetzler@umd.edu\\}

\begin{document} 
\maketitle

\begin{abstract}
ComPair is a prototype gamma-ray telescope for the development of key technologies for next-generation gamma-ray detectors consisting of four subsystems: a 10-layer double-sided silicon strip detector tracker, a cadmium zinc telluride calorimeter, a cesium iodide calorimeter, and a plastic anti-coincidence detector (ACD). The ACD acts as an active shield to veto charged particle events and consists of 5 plastic scintillating panels. ComPair was launched as a balloon payload from Ft. Sumner, New Mexico and completed a 6-hour flight on August 27, 2023. Here we detail the design and calibration of the ComPair ACD, and report on the ACD’s veto efficiency and other performance metrics during the ComPair flight.
\end{abstract}


\section{INTRODUCTION}
\label{sec:intro}  

\begin{figure}
    \centering
    \includegraphics[width=0.5\linewidth]{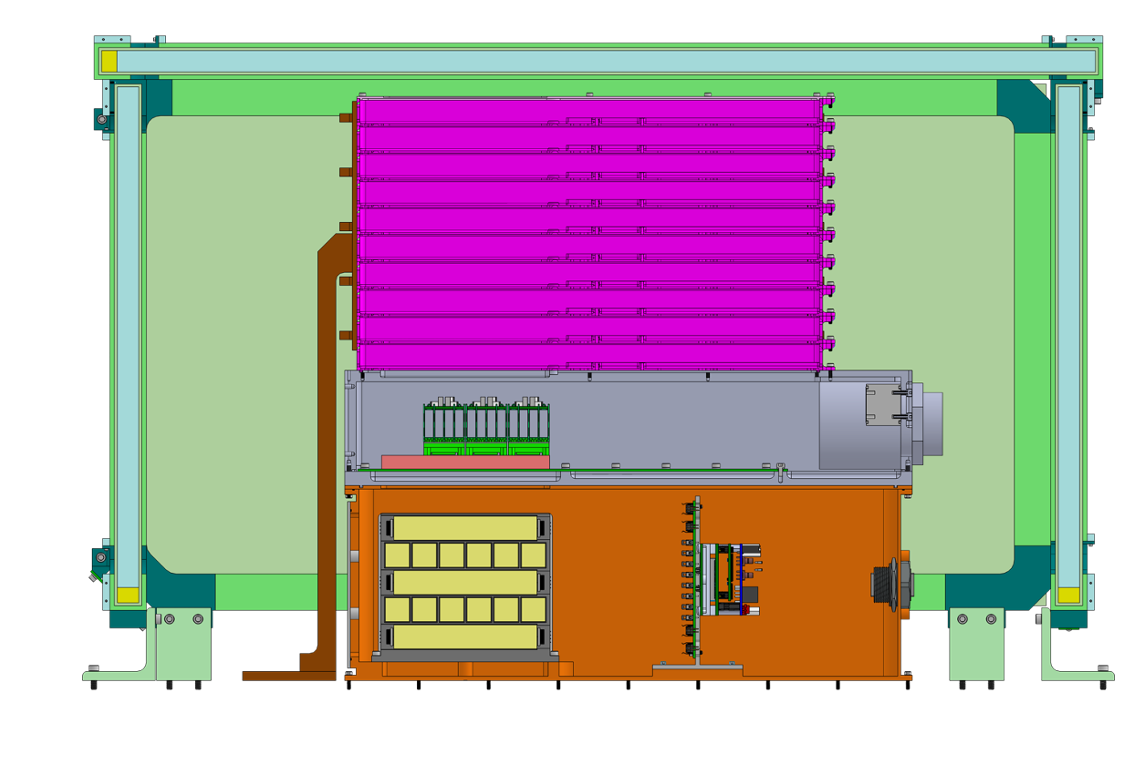}
    \caption{ComPair consists of a double-sided silicon strip detector tracker (purple), a cadmium zinc telluride calorimeter (gray), a CsI calorimeter (orange), and a plastic scintillator ACD (green).}
    \label{fig:CAD}
\end{figure}

Charged particles are the largest source of background in the MeV gamma-ray regime. As such, gamma-ray telescopes must reject charged particle events through either active shielding or pulse-shape discrimination. This proceeding reports on the design and performance of the anti-coincidence detector (ACD) of ComPair, the Compton-Pair Telescope. \cite{Valverde_2023,2022SPIE12181E..2GS} ComPair is a prototype of the All-sky Medium Energy Gamma ray Observatory (AMEGO), \cite{mcenery2019allsky,2020SPIE11444E..31K} a probe-class mission concept to fill the gap in gamma-ray sensitivity between 100 keV and 1 GeV. ComPair consists of a 10-layer double-sided silicon strip detector tracker,\cite{2020SPIE11444E..34G,griffin2019development,kirschner} a cadmium zinc telluride (CZT) virtual Frisch grid calorimeter,\cite{Hays:2020onp,Orlando_2022,Sasaki:2023O8} a cesium iodide calorimeter (CsI),\cite{2023ITNS...70.2329S,2019LPICo2135.5016W,shy} and a plastic scintillator ACD. A model of ComPair is presented in Figure \ref{fig:CAD}. ComPair completed a short duration high altitude balloon flight on August 27, 2023.\cite{smith}

The ComPair ACD is an active shield to veto charged particle events. The Fermi Large Area Telescope (LAT)\cite{Atwood_2009} also uses a plastic scintillator ACD\cite{LAT_ACD,MOISEEV2007372}, and the ComPair design leverages this heritage. The ComPair ACD design and integration procedure are described in Section \ref{sec:design}. The data acquisition, calibration, and analysis pipelines are described in Section \ref{sec:pipeline}. Pre-integration testing is covered in Section \ref{sec:pre-integration}. Two pre-flight validation runs are discussed in Section \ref{sec:pre-flight}. The results from the balloon flight are reported on in Section \ref{sec:flight}.

\section{The ACD Design}\label{sec:design}

\begin{figure}
    \centering
    \begin{subfigure}{0.35\textwidth}
        \includegraphics[width=\linewidth]{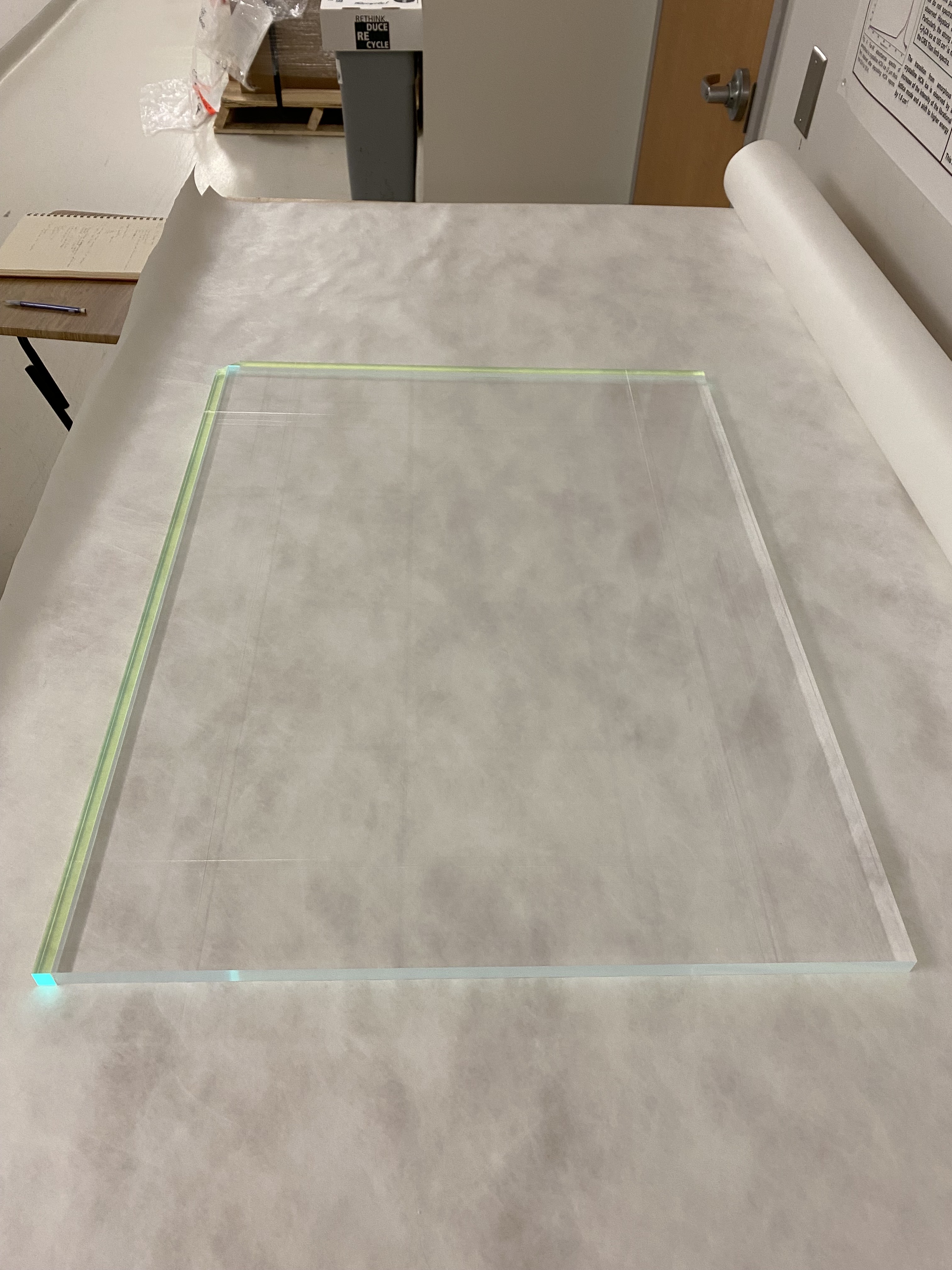}
        \caption{}
        \label{fig:panelUnwrapped}
    \end{subfigure}
    \hfill
    \begin{subfigure}{0.55\textwidth}
        \includegraphics[width=\linewidth]{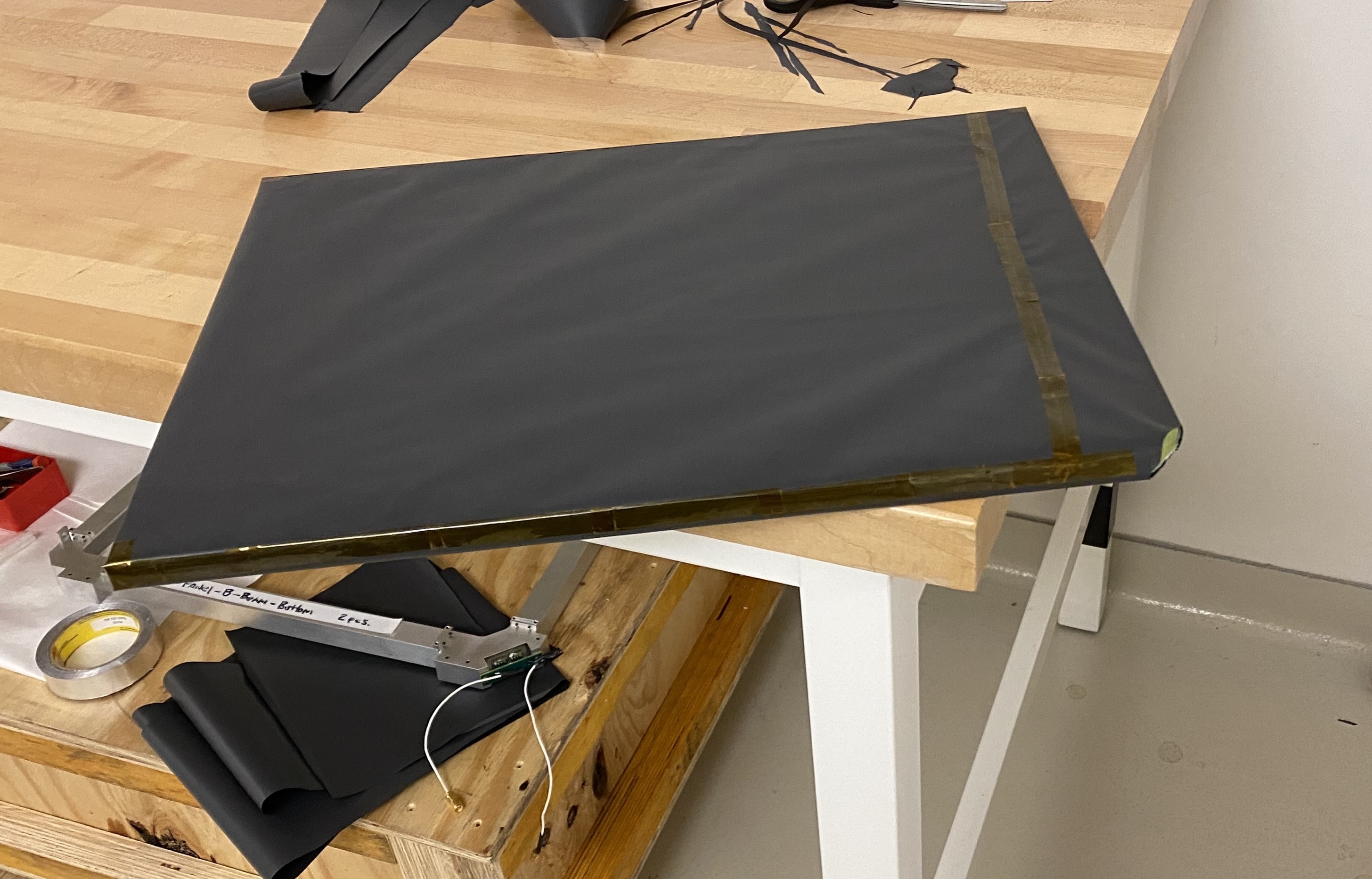}
        \caption{}
        \label{fig:panelTedlar}
    \end{subfigure}
    \hfill
    \begin{subfigure}{0.35\textwidth}
        \includegraphics[width=\linewidth]{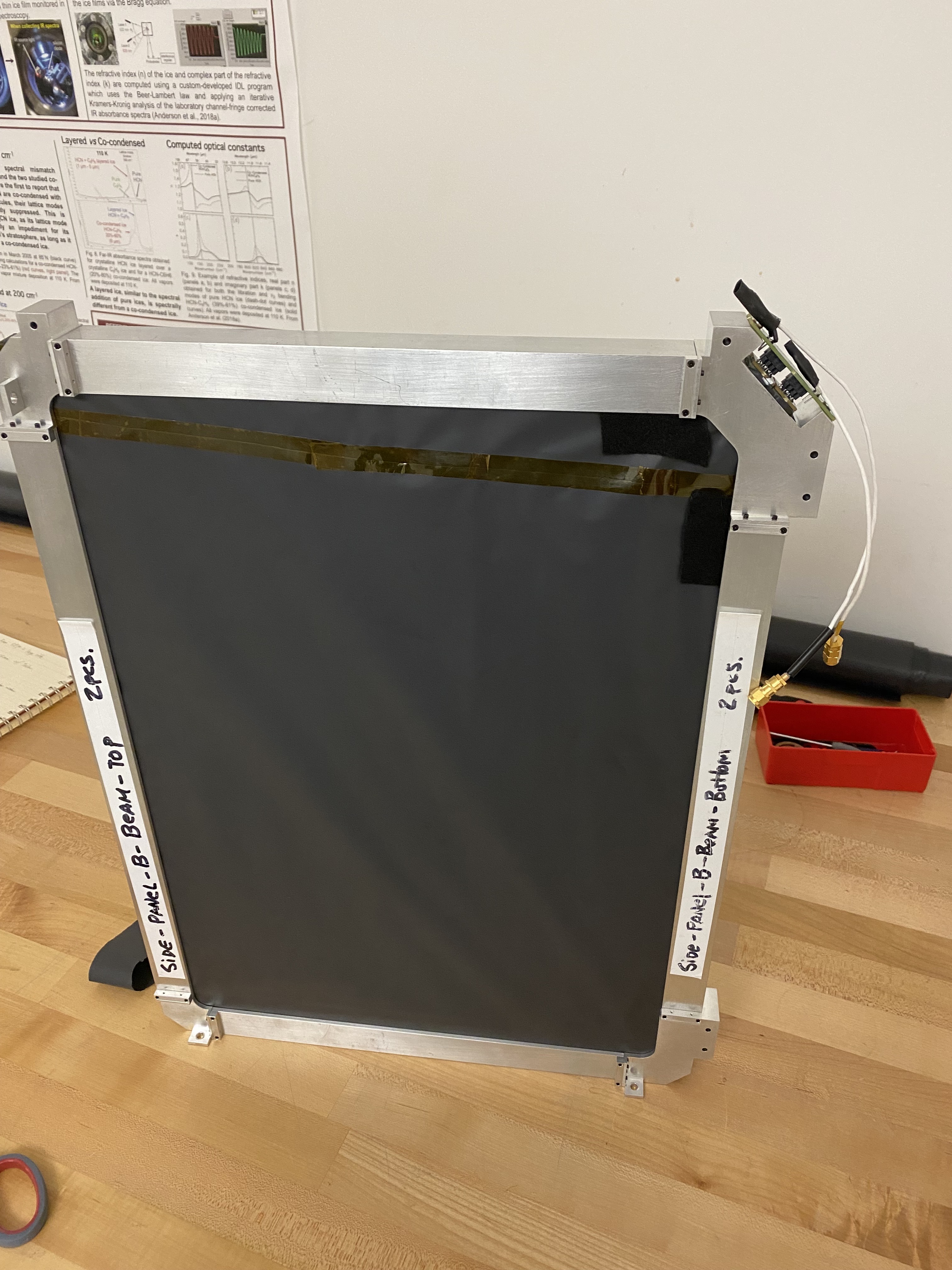}
        \caption{}
        \label{fig:panelFrame}
    \end{subfigure}
    \hfill
    \begin{subfigure}{0.55\textwidth}
        \includegraphics[width=\linewidth]{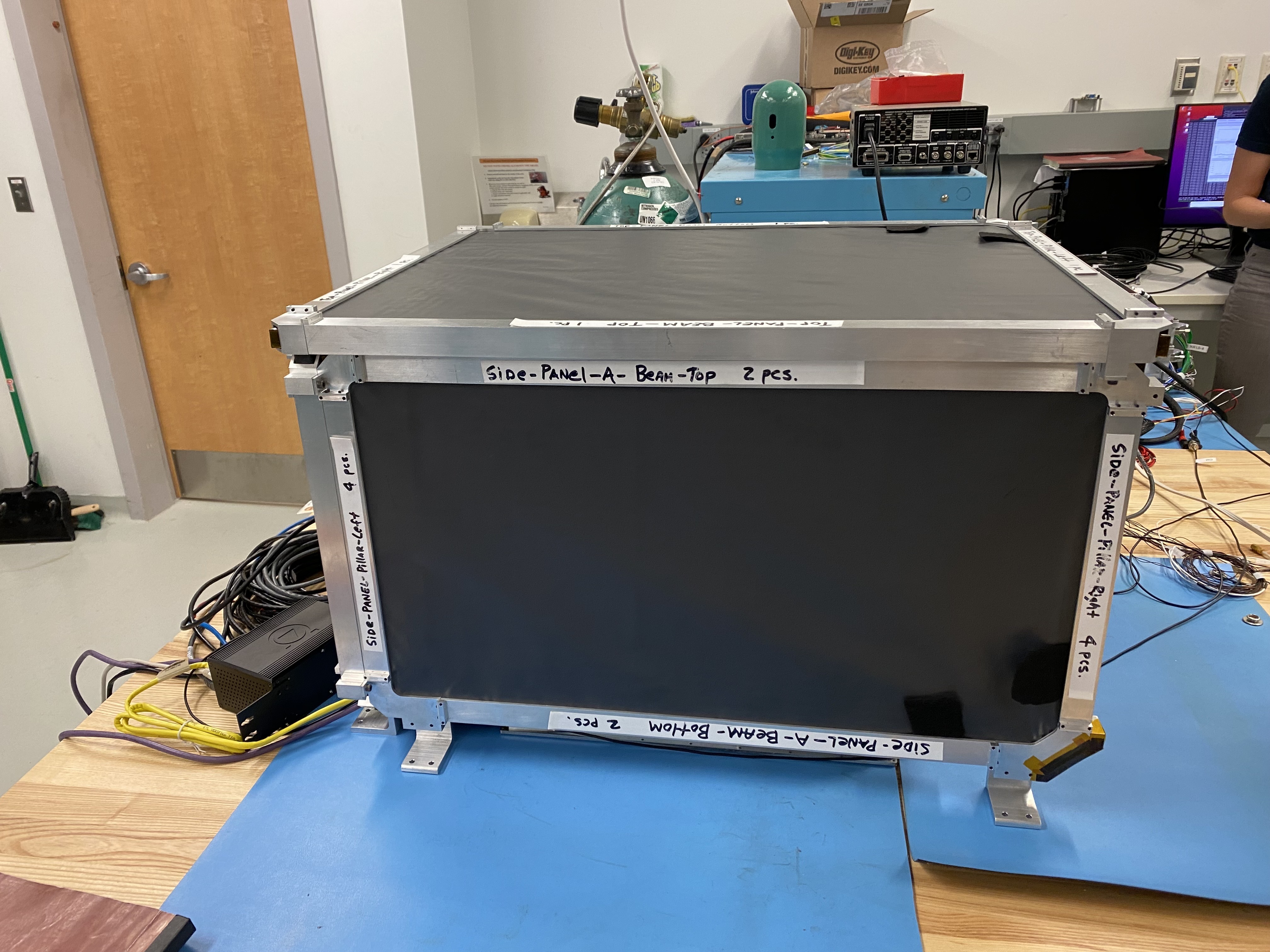}
        \caption{}
        \label{fig:panelAssembled}
    \end{subfigure}
    \hfill
    \caption{\ref{fig:panelUnwrapped}: Two WLSs are pressed against each panel on the top and left. The WLSs guide the light to SiPMs (not shown). In the top left corner of the assembly, beveled edges of each WLS will face a SiPM. Tyvek (in the background) will be wrapped around the module as a diffuse reflector. \ref{fig:panelTedlar}: An ACD panel is wrapped with 2 layers of Tyvek and 3 layers of Tedlar. The SiPMs (not shown) will be pressed against beveled faces of the WLSs in the bottom right of the figure. The wrapping is held together with Kapton tape. \ref{fig:panelFrame}: An aluminum frame surrounds the panel assembly and holds the SiPMs (top right) against the WLS bars. Black felt is folded around the side of the Tyvek-wrapped WLSs to maintain light-tightness while allowing air to escape the wrapping during flight. \ref{fig:panelAssembled}: 5 panels comprise the ACD assembly. Each corner with SiPMs is covered in an additional layer of Tedlar for light-tighting.}
\end{figure}

The ComPair ACD consists of five modules surrounding the four sides and top of the telescope. One module consists of a polyvinyltoluene scintillating panel (Eljen EJ 208),\cite{Panels} coupled to two Eljen EJ 280 wavelength shifting bars (WLSs) that absorb the scintillation light and re-emit isotropically at 490 nm.\cite{WLSs} Each WLS is read out by a 2 $\times$ 2 Onsemi C-Series silicon photomultiplier array (SiPM).\cite{OnSemi,BUZHAN200348} Each module is wrapped with two layers of Tyvek as a diffuse reflector and three layers of Tedlar for light-tighting. The module is held together mechanically by an aluminum frame. Figures \ref{fig:panelUnwrapped}-\ref{fig:panelAssembled} show the stages of the ACD assembly, beginning with an unwrapped panel and WLSs on Tyvek, then a fully wrapped module, a module in its aluminum frame, and finally the fully assembled ACD. The light-tightness of each panel was tested with normal ambient lighting. The slightest leak renders the ACD unable to record data due to such a high photon count rate. Any light leaks were patched with aluminum tape.

The dimensions of the panels are listed in Table \ref{Tab:ACD Dimensions}. The WLSs are the same length and width as the edges of the panels and are 10.6mm thick. Each WLS is beveled at a $45^{\circ}$ angle on one end for the SiPM face. In addition, we procured one more panel for testing purposes that was not incorporated into the final ACD; this extra panel is later called the ``test panel.''

\begin{table}[ht]
\caption{The ACD has one top panel, and two each of the side panel sizes. Each panel has two WLSs of the same length and width. The WLSs are 10.6 mm thick.} 
\begin{center}       
\begin{tabular}{|l|l l l|} 
\hline
\textbf{Part Name} & \multicolumn{3}{|l|}{\textbf{Panel  Dimensions (mm)}}\\
\hline
Top Panel & 683 &$\times$ 523 &$\times$ 15  \\
\hline
Large Side Panel & 636 &$\times$ 350 &$\times$ 15 \\
\hline
Small Side Panel & 476 &$\times$ 350 &$\times$ 15 \\
\hline 
\end{tabular}
\label{Tab:ACD Dimensions}
\end{center}
\end{table}

Due to the unique light collection scheme, where each panel's light is collected in one corner, it is important to have long attenuation lengths in the plastic scintillator. Compared to other Eljen plastic scintillators, EJ 208 has the longest reported attenuation length of 4 m. The optical and mechanical characteristics of EJ 208 and Eljen's alternative materials are shown in Table \ref{Tab:plastic scintillators}. Each Eljen plastic scintillator in this series has a reported refractive index of 1.58, a coefficient of linear expansion of $7.8 $x$ 10^{-5}$ below 67 \textdegree C, no temperature-dependent light output change between -60 \textdegree C and  20 \textdegree C, and is vacuum compatible. Due to its similarity with EJ 208, EJ 200 would also have been appropriate for the ACD.

The operating principle for the WLSs is that the scintillating light will be absorbed and re-emitted in all directions, allowing some of the re-emitted light to propagate directly to the SiPMs no matter where the initial interaction occured in the panel. The WLSs were chosen to maximize both absorption of the scintillated light and emission of light at wavelengths where the SiPM light collection efficiency is high. EJ 280 maximizes absorption at 427 nm (well matched to EJ 208's emission at 435 nm) and maximizes emission at 490 nm. The index of refraction and coefficient of linear expansion for the WLSs match those of the scintillating panels.

The Onsemi C-Series SiPMs (ArrayC-60035-4P-BGA) were chosen for high gain and sensitivity to visible light at 490 nm, as well as improved quantum efficiency compared to photomultiplier tubes. Each SiPM has a 2 $\times$ 2 array of 6 mm $\times$ 6 mm active areas with 35 $\mu$m microcells. The listed breakdown voltage is between 24.2 and 24.7 V, and they were biased at 28 V.

\begin{table}[ht]
\caption{Eljen provides optical characteristics for its plastic scintillators\cite{Panels}. The ComPair ACD uses EJ 208 for its long attenuation length.} 
\begin{center}       
\begin{tabular}{|l|l|l|l|l|} 
\hline
\textbf{Properties} & \textbf{EJ 200} & \textbf{EJ 204} & \textbf{EJ 208} & \textbf{EJ 212}  \\
\hline
Scintillation Efficiency (Photons/1 MeV $e^{-}$) & 10000 & 10400 & 9200 & 10000  \\
\hline
Wavelength of Maximum Emission (nm) & 425 & 408 & 435 & 250 \\
\hline
Light Attenuation Length (cm) &	380 &	160 &	400 &	250 \\
\hline
Rise Time (ns) &	0.9 &	0.7 &   1.0 &	0.9 \\
\hline 
Decay Time (ns) &	2.1 &	1.8 &	3.3 &	2.4 \\
\hline
Pulse Width, FWHM (ns)  &	2.5 &	2.2 &	4.2 &	2.7 \\
\hline
\end{tabular}
\label{Tab:plastic scintillators}
\end{center}
\end{table}

\begin{figure}
    \centering
    \includegraphics[width=0.5\linewidth]{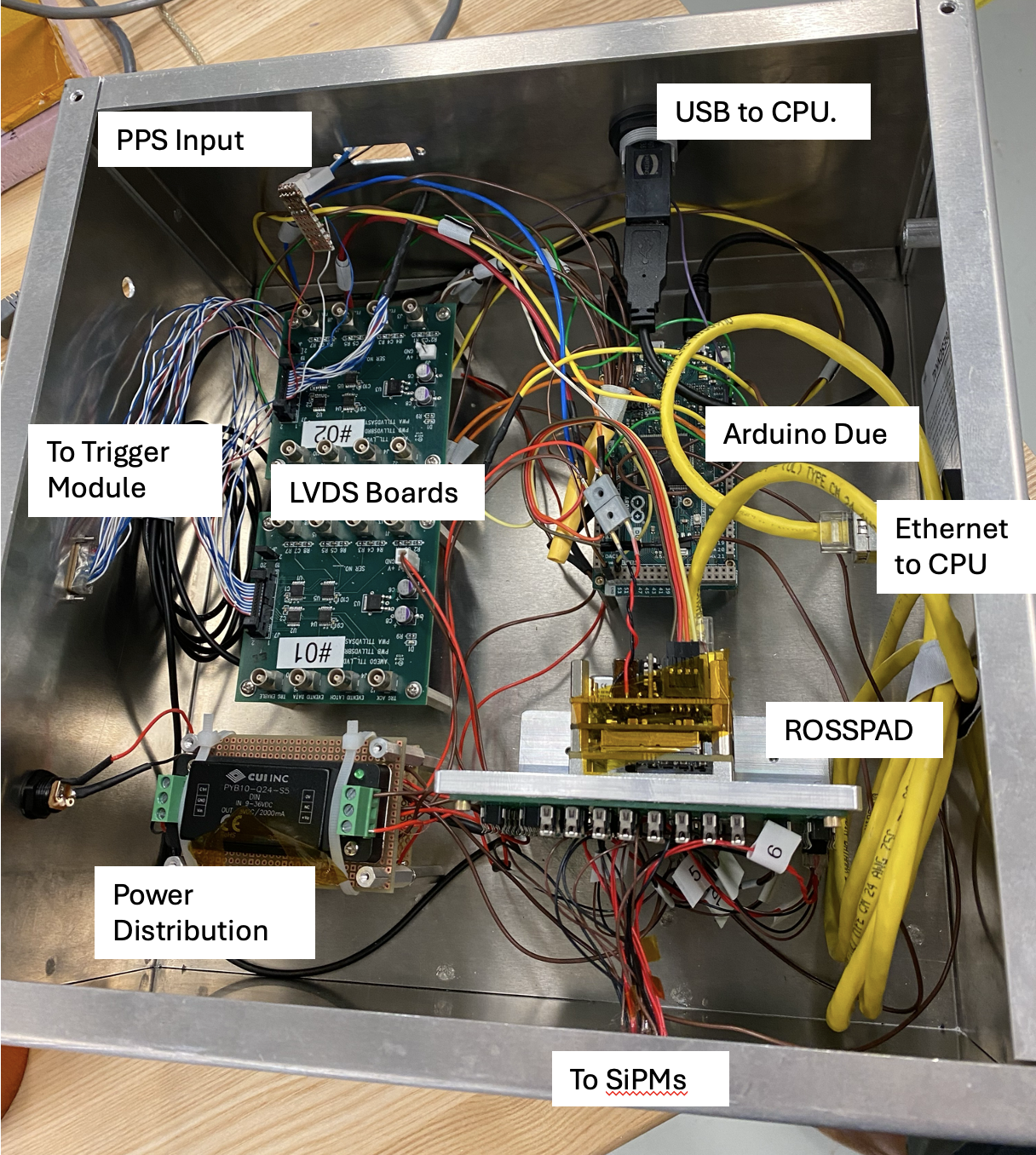}
    \caption{A power distribution unit supplies 5 V to the rest of the components in the box. Inputs from the Trigger Module and PPS distribution are transformed to the correct voltages by the LVDS boards for the Arduino and ROSSPAD. The ComPair CPU communicates with the Arduino Due over USB, and with the ROSSPAD over ethernet.}
    \label{fig:electronics}
\end{figure}

The SiPMs are read out by an IDEAS ROSSPAD, a commercially available system designed to read out up to 64 SiPMs.\cite{ROSSPAD} The ROSSPAD includes 4 SIPHRA ASICs,\cite{SIPHRA} which perform on-chip pulse shaping, digitization, triggering, and data transmission to the ComPair flight computer. When the signal from any individual SiPM is above the trigger threshold, all channels are read out and recorded. This scheme means that data is recorded in panels that do not have an interaction, effectively sampling the noise floor. This affects the side panels more than the top panel, and the consequences will be discussed later in Section \ref{sec:flight}. An Arduino Due is used to correlate the timestamps in the ROSSPAD and Trigger Module, discussed further in Section \ref{sec:pipeline}. Figure \ref{fig:electronics} shows a photograph of the ACD electronics and all its components.

\section{ACD Data Acquisition and Analysis Pipeline}\label{sec:pipeline}

\subsection{Event Alignment}

\begin{figure}
    \centering
    \includegraphics[width=0.7\linewidth]{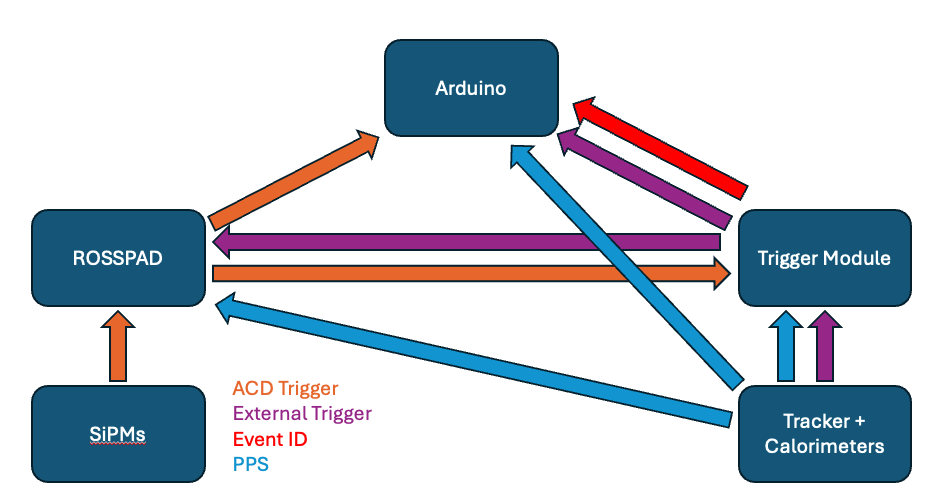}
    \caption{The ACD event alignment requires matching ACD triggers that originate in the ACD itself and external triggers, which originate in the other ComPair detectors. In addition to the triggers, the Arduino collects the UTC times of the PPS signals and unique Event IDs for each external trigger.}
    \label{fig:electronics flowchart}
\end{figure}

The ACD readout electronics consist of an IDEAS ROSSPAD, an Arduino Due, and a low-voltage differential signaling interface (LVDS) to the ComPair Trigger Module.\cite{2020SPIE11444E..6AS} The Arduino is necessary, because the ROSSPAD's firmware cannot be changed to record everything needed for event alignment. The flow of information is shown in Figure \ref{fig:electronics flowchart}, and a photograph of the ACD electronics are shown in Figure \ref{fig:electronics}. The ACD tabulates data from either ACD triggers, which originate in the ROSSPAD when any SiPM readout rises above a predetermined threshold that corresponds to about 3 MeV, or external triggers, which originate in the Trigger Module when specific combinations of the tracker or calorimeters have their own internal triggers in coincidence. The ComPair tracker has individual trigger primitives for each side of its 10 layers, while the calorimeters and the ACD have only one trigger primitive each. The trigger conditions used for ComPair are:

\begin{enumerate}
    \item Any 2 tracker primitives
    \item A single tracker primitive and the CZT calorimeter
    \item A single tracker primitive and the CsI calorimeter
    \item The CZT calorimeter and the CsI calorimeter
\end{enumerate}

When one of the 4 trigger conditions are met, the Trigger Module assigns a unique Event ID and issues an external trigger to all of the detector subsystems. The ROSSPAD creates an event packet in its log without reading the SiPMs out, and the Arduino records both the external trigger and the value of the Event ID. At the beginning of data collection, the Trigger Module also sends 10 external triggers to the ACD to align the beginning of the run in the ROSSPAD and Arduino logs. Similarly, a pulse-per-second (PPS) signal is sent to the Trigger Module, the ACD ROSSPAD, and the ACD Arduino from the CsI calorimeter. The ROSSPAD and Arduino each record the arrival time of the PPS signal in terms of their own internal counters. The Arduino also records the UTC time of the PPS signal's arrival. This PPS signal is important, because the ROSSPAD and Arduino only have internal counters, and without the PPS signal there is no way to align the ROSSPAD and Arduino logs in time and ultimately assign Event IDs to interactions in the ACD. Finally, the ROSSPAD sends a signal to the Arduino and the Trigger Module for each ACD trigger.

Event alignment occurs after data collection is complete. The first step is to align the ROSSPAD counter with the Arduino counter using the 10 external triggers at the beginning of the run and the PPS signals. The times are cast from the arbitrary internal counters on both the ROSSPAD and Arduino to UTC times. Then, ACD triggers and external triggers are matched between the ROSSPAD and Arduino logs using the UTC times. Finally, ACD triggers and external triggers that are separated by more than 20 $\mu$s are rejected, since these cannot be reliably associated with events in other subsystems.

Unfortunately, both the ACD and CsI Arduinos stopped recording data early in the balloon flight due to an as-yet undetermined failure. As such, the data analysis pipeline now uses redundant data from the Trigger Module log. The process is largely the same as before, but now the ROSSPAD internal clock times are transformed only to the Trigger Module internal clock time as opposed to UTC times. Neither the ROSSPAD nor the Trigger Module record the UTC time of the PPS signals, only that the signals are received. This is expanded upon in Section \ref{sec:flight}. Whether or not the Arduino is functioning properly, each of the ACD triggers that occur within 20 $\mu$s of an external trigger is assigned an Event ID and used for analysis with the other subsystems.

\subsection{Simulation Pipeline}

We use the Excel-based Program for calculating Atmospheric Cosmic-ray Spectrum\cite{10.1371/journal.pone.0160390} particle background model to estimate the expected flux for each particle species during flight and ground calibrations. High fidelity simulations are necessary to understand and calibrate the ACD. We consider photons, electrons, positrons, protons, neutrons, and muons. A Monte Carlo simulation is performed with Cosima, the Cosmic Simulator of MEGAlib\cite{2006NewAR..50..629Z} to calculate the energy deposits in each detector subsystem. 

The output from Cosima is passed through the ComPair Detector Effects Engine (DEE), an internally-developed python package that applies instrumental effects not included in MEGAlib and transforms the data to resemble the raw output from the instruments. The output of the DEE is then passed through the same data analysis pipeline as the real data to arrive at realistic simulated data.

\subsection{Calibration}

The standard energy calibration process for gamma-ray detectors uses multiple radioactive sources with known emission lines. However, the ACD is insensitive to gamma-ray emission by design and requires a different calibration method. A linear calibration can be achieved using only a single run by matching the measured spectrum due to the ambient particle background from the ROSSPAD analog-to-digital converter (ADC) with simulations of the expected energy deposit spectra. For calibration, we do not apply the DEE to the simulations. We also neglect nonlinearities in the response at larger energy deposits, since the primary function of the ACD is to veto charged particles, and we do not expect the veto efficiency to decrease with larger energy deposits.

\begin{figure}
    \centering
    \includegraphics[width=0.5\linewidth]{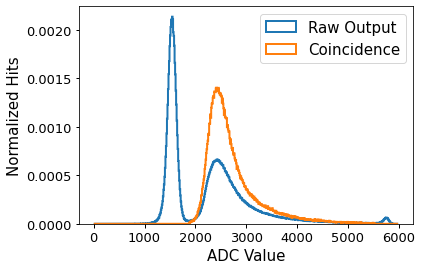}
    \caption{Requiring coincidence between all 5 ACD panels during calibration suppresses the noise. For example, we show the raw ADC spectra for one of the SiPMs on the largest ACD panel. The raw output on this channel is shown in blue, and the spectrum when both SiPM channels on all 5 panels are triggered is shown in orange.}
    \label{fig:noise suppression}
\end{figure}

For the calibration run, the 5 ACD panels are stacked with 5 cm spacers between each panel, and we require all 5 panels to be triggered. This selection significantly suppresses the noise pedestal in the measured spectra. The noise suppression for the SiPM facing the shorter edge of the top panel is demonstrated in Figure \ref{fig:noise suppression}. Similarly, the simulated energy deposits assume 5 panels stacked and require an event to interact in all panels.

For each panel, the measured and simulated spectra are fit by a Landau distribution\cite{Landau:1944fvs}, shown in Equation \ref{Landau Equation}. The Landau distribution describes the energy deposited in a material by charged particle ionization, and is fit with 2 parameters, $\mu$ and $\eta$, where $\mu$ is the location of the peak and $\eta$ represents the width of the distribution. 

\begin{equation}
    p(x,\mu,\eta) = \frac{1}{\pi \eta} \int_0^\infty dt\, 
 e^{-t} \,cos \left(t\,\left(\frac{x-\mu}{\eta}\right)+\frac{2t}{\pi} \,log\left(\frac{t}{\eta}\right)\right)
 \label{Landau Equation}
\end{equation}

\begin{figure}
    \centering
    \includegraphics[width=0.5\linewidth]{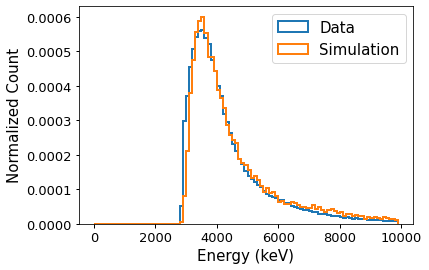}
    \caption{Calibrated energy deposits as measured by the SiPM facing the shorter edge of the largest ACD panel are shown in blue, and the simulated energy deposits from Cosima for the same panel are shown in orange.}
    \label{fig:calibrated spectra}
\end{figure}

We determine the conversion from ADC value to energy using the points $\mu - \eta$ and $\mu$ from the Landau fit in ADC space from the measured spectra and the same points from the Landau fit in energy space from the simulated spectra. Figure \ref{fig:calibrated spectra} shows the calibrated and simulated spectra for the the same channel as in Figure \ref{fig:noise suppression}. We also estimate the ACD energy resolution using the fit uncertainties from $\mu$ and $\eta$ for each of the measured and simulated spectra. The resolution for each channel is shown in Table \ref{Tab:Uncertainties}. These energy resolution values are in line with other tests of plastic scintillator energy resolution of 10-15\% standard deviation \cite{mengesha2015plastic}. 

For the analysis in later sections, one SiPM each for two of the side panels are excluded due to issues with the gain on those two channels during pre-flight validation and the balloon flight itself. One of the SiPMs facing the horizontal edge of a small side panel had poor light collection efficiency to the point that the Landau distribution was not separated from the noise pedestal. The other unused SiPM faced the horizontal edge of a large side panel and will be discussed in Section \ref{sec:flight}.

\begin{table}[ht]
\caption{The uncertainties for the fit parameters, $\mu$ and $\eta$, can be used to estimate the energy resolution of the ACD. The energy resolution estimates are consistent with other tests of plastic scintillators. Panels 1 and 3 are small side panels, 2 and 4 are large side panels. Panel 5 is the top panel. For the side panels, SiPM 1 faces the horizontal WLS and SiPM 2 faces the vertical WLS. For the top panel, SiPM 1 faces the shorter WLS, and SiPM 2 faces the longer WLS. SiPM 1 on panels 1 and 2 were not used for the analysis in Sections \ref{sec:pre-flight} and \ref{sec:flight} due to gain-related issues.} 
\begin{center}       
\begin{tabular}{|c|l|l|l|l|l|} 
\hline
 & \multicolumn{2}{c|}{Data} & \multicolumn{2}{c|}{Simulation} & \\
\hline
Panel, SiPM Number & $\sigma_\mu$ (ADC) & $\sigma_\eta$ (ADC) & $\sigma_\mu$ (keV) & $\sigma_\eta$ (keV) & $\sigma_E$ @ 5 MeV \\
\hline
 1, 1 & 1.9 & 1.2 & 3.3 & 2.4 & 704 keV\\
\hline
 1, 2 & 3.9 & 1.3 & 3.3 & 2.4 & 327 keV\\
\hline
 2, 1 & 3.3 & 1.3 & 4.0 & 2.9 & 339 keV\\
\hline
 2, 2 & 7.6 & 1.3 & 4.0 & 2.9 & 663 keV\\
\hline
 3, 1 & 5.5 & 1.9 & 2.1 & 1.6 & 398 keV\\
\hline
 3, 2 & 4.6 & 2.0 & 2.1 & 1.6 & 362 keV\\
\hline
 4, 1 & 4.0 & 1.3 & 4.7 & 3.5 & 416 keV\\
\hline
 4, 2 & 3.1 & 1.3 & 4.7 & 3.5 & 355 keV\\
\hline
 5, 1 & 2.8 & 1.1 & 4.0 & 3.0 & 354 keV\\
\hline
 5, 2 & 4.5 & 1.2 & 4.0 & 3.0 & 451 keV\\
\hline
\end{tabular}
\label{Tab:Uncertainties}
\end{center}
\end{table}

\section{Pre-integration Testing} \label{sec:pre-integration}

\begin{figure}
    \centering
    \includegraphics[width=0.5\linewidth]{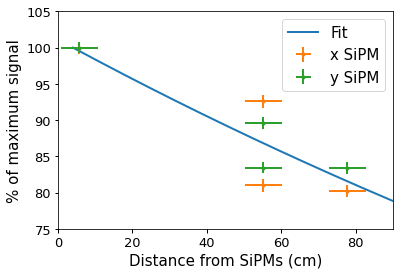}
    \caption{The light collection efficiency was measured using a muon telescope that selected charged particles passing through the four corners of the test panel. Defining the corner nearest the SiPMs as the maximum signal at 100\%, we measured the shift in $\mu$ with respect to the value at this nearest corner. The results are consistent with the attenuation length quoted by Eljen. The x SiPM result at 100\% is directly behind the point for the y SiPM, which is why it cannot be seen.}
    \label{fig:panel loss}
\end{figure}

To assess light collection efficiency across a panel, measurements were performed using the test panel, mentioned in Section \ref{sec:design}. The test panel measures 55 $\times$ 55 $\times$ 1.5 cm$^3$ and is wrapped similarly to the panels used for the ACD. This test also utilized two small plastic scintillating paddles, each coupled to a single SiPM to form a muon telescope. By placing one paddle above and one below a spot on the test panel and selecting events that trigger both paddles, we isolated events that pass through a specific region of the test panel. This was performed four times, once in each corner of the test panel in order to gauge the signal lost as light travels from the far corner of the panel to the SiPM readout. For the two SiPMs coupled to the corner of the test panel, a baseline was subtracted from the raw ADC spectra, which was then fit by a Landau distribution. We measured the loss across the panel by comparing the change in $\mu$ to the distance from each SiPM. The results are shown in Figure \ref{fig:panel loss}. These losses were fit with an exponential decay function with an attenuation length of 362 $\pm$ 39 cm, which is reported with uncertainty at the 1 $\sigma$ level and which is consistent with the 400 cm quoted by Eljen as in Table \ref{Tab:plastic scintillators}. The largest ACD panel is 86 cm diagonally, so it is expected to have a 20\% reduction in signal at the corner farthest from the SiPM readout. This level light collection loss is acceptable for the ComPair balloon flight.

\section{Pre-Flight Validation}\label{sec:pre-flight}

To ensure that ComPair was working properly before flight, it was tested with several runs with and without radioactive sources during the Fort Sumner campaign. The full instrument calibration required measurements with radioactive sources; however, measurements of the ambient background radiation and with a Cs-137 were used to evaluate the ACD performance.

\subsection{Ambient Particle Background}

A one hour run without any additional radioactive sources is the simplest test to confirm the ACDs operation. The key indicator to evaluate the ACD performance is the veto rate, defined as number of charged particle events that triggered the ACD divided by the total number of charged particle events. However, estimating the number of charged particle events for the denominator is difficult, because by definition the ACD cannot be used to make this determination.

\begin{figure}
    \centering
    \includegraphics[width=0.9\linewidth]{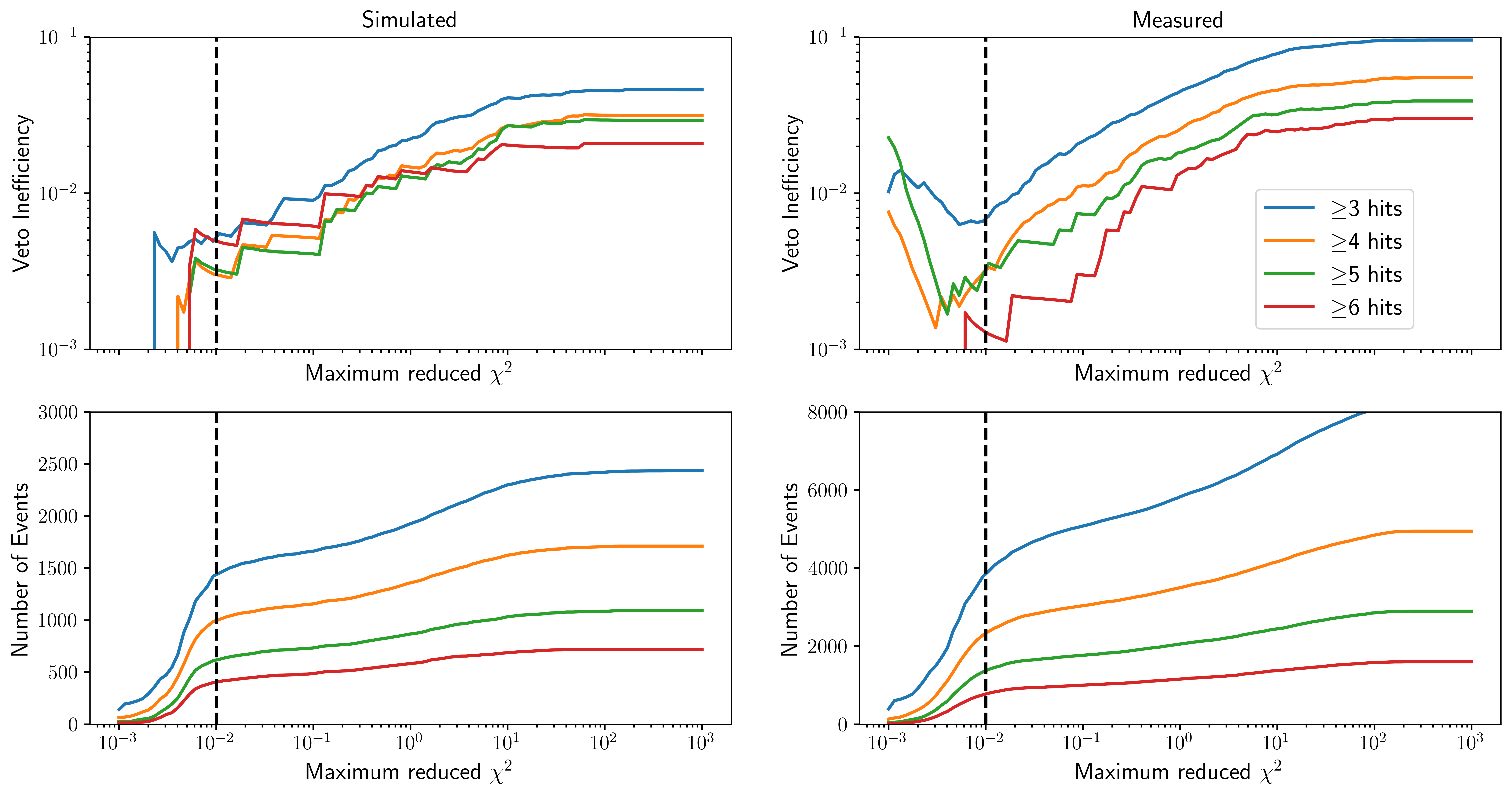}
    \caption{We compared the veto inefficiency for both a simulated and measured ambient particle background in Fort Sumner, New Mexico in August 2023. Each solid line corresponds to a minimum number of tracker hits required per event. The vertical, dashed line represents the reduced $\chi^2$ value used to determine the ACD veto rate. The bottom row shows the number of events meeting both constraints. The simulated data set covers 30 minutes, while the measured data set covers 1 hour.}
    \label{fig:run286Veto}
\end{figure}

To determine the true number of incident particle events data from the ComPair Tracker is used. The ComPair Tracker detects both gamma rays and charged particles, but charged particles should follow straight paths through the tracker. On the other hand, gamma rays will either Compton scatter or produce electron/positron pairs. Both of these gamma-ray interactions are unlikely to follow a single straight line. However unlikely, the Compton scattered electron could continue in a roughly straight line. Also, a pair-produced electron or positron may escape the detector such that its companion looks like a single tracked charged particle. To select a group of events with the highest likelihood of being charged particles, we placed constraints on the minimum number of interactions in the tracker as well as the maximum reduced $\chi^2$ value of the track fit to a straight line. The result is reported as a veto inefficiency, which is one minus the veto rate. Figure \ref{fig:run286Veto} shows the veto inefficiency and the number of qualifying events for minimum numbers of hits and maximum reduced $\chi^2$ for the particle track in the tracker for both simulated and measured data sets. Considering events with at least 4 hits and reduced $\chi^2 < 0.01$, the ACD had both measured and simulated veto rates of 99.7\%. A reduced $\chi^2 < 1$ is considered over-fit, and the source of over-fitting for this analysis is due to using conservative values for the tracker position uncertainties.

\begin{figure}
    \centering
    \includegraphics[width=0.9\linewidth]{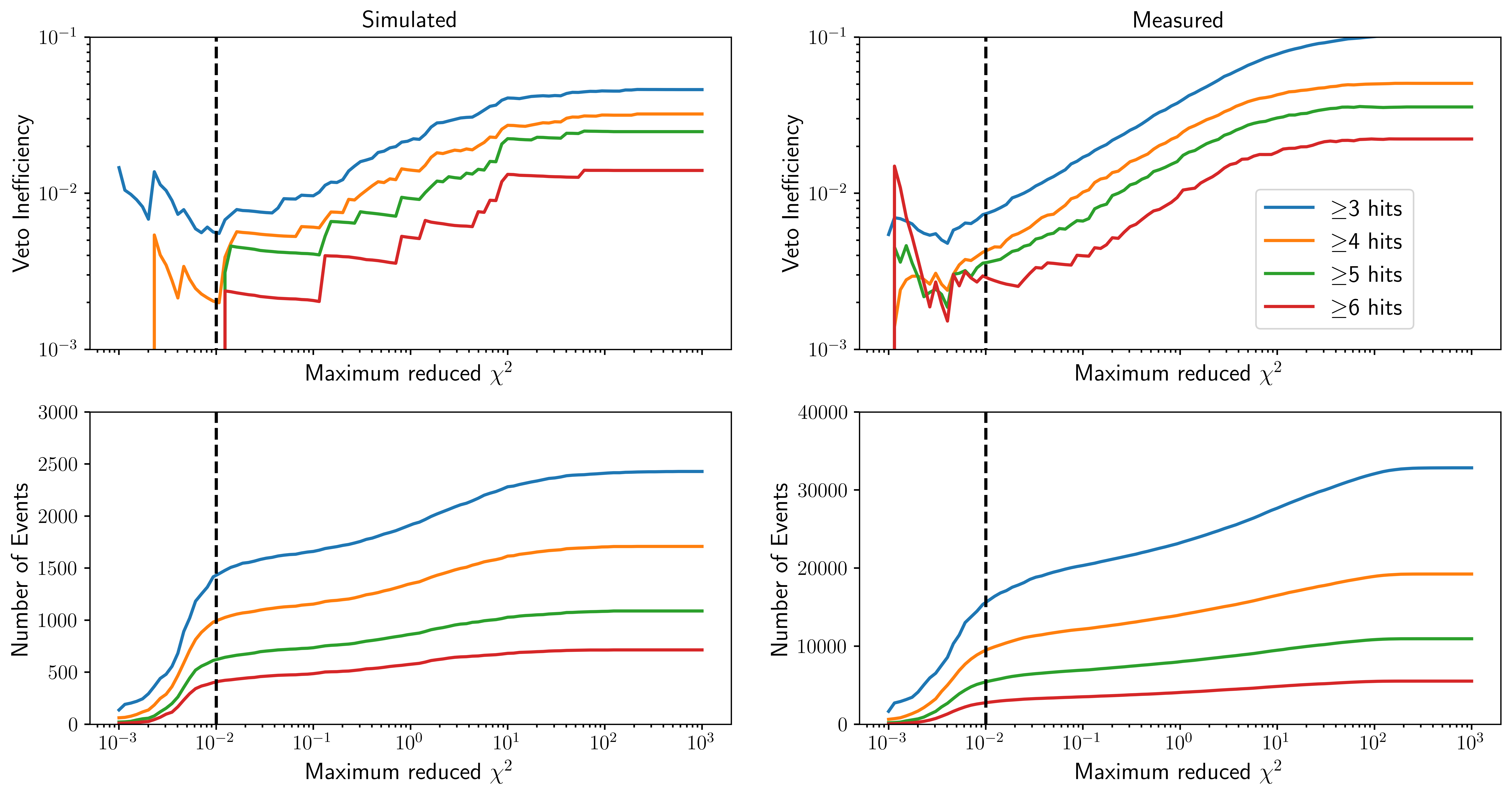}
    \caption{The veto inefficiency for both a simulated and measured Cs-137 source in Fort Sumner, New Mexico in August 2023 was calculated in an identical manner to the ambient particle background. Each solid line corresponds to a minimum number of tracker hits required per event. The vertical, dashed line represents the reduced $\chi^2$ value used to determine the ACD veto rate. The bottom row shows the number of events meeting both constraints. The simulated data set covers 30 minutes, while the measured data set covers 6.5 hours.}
    \label{fig:run300Veto}
\end{figure}

\subsection{Cs-137 Source}

A second useful diagnostic run used a Cs-137 source, which emits gamma rays at 661.7 keV. This source was placed 2 m above ComPair for 6.5 hours. The purpose of this test was to ensure that the event selection used to identify charged particles with the tracker are robust when gamma rays are added to the environment. The veto rates were calculated in an identical manner to the ambient particle background run. The results from both measured and simulated datasets are shown in Figure \ref{fig:run300Veto}. Using the same constraints as the ambient background, the ACD had measured and simulated veto rates of 99.6\% and 99.8\%, respectively. Since the veto rate did not differ appreciably from the ambient particle background run, we are confident that the event selection effectively discriminates between gamma rays and charged particles and that the ACD was not vetoing gamma rays.

\begin{figure}
    \centering
    \includegraphics[width=0.5\linewidth]{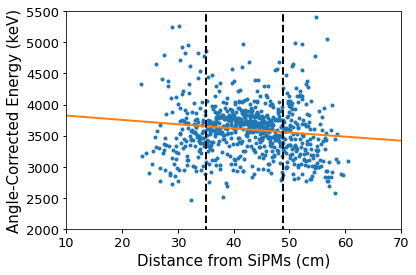}
    \caption{The average angle-weighted energy deposit in the ACD top panel in 1 $\times$ 1 cm bins for the Cs-137 source. The black dashed lines are the horizontal distance of the closest and furthest corners of the tracker detector stack to the top panel's SiPMs. To find the attenuation length, we fit the data to an exponential decay function (orange), which looks linear due to the long attenuation length compared to the size of the panel. The attenuation length was found to be 537 $\pm$ 169 cm.}
    \label{fig:run300LossScatter}
\end{figure}

\begin{figure}
    \centering
    \begin{subfigure}{0.49\textwidth}
        \includegraphics[width=\linewidth]{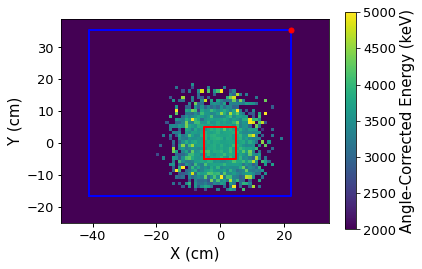}
        \caption{}
        \label{fig:run300HeatMap}
    \end{subfigure}
    \hfill
    \begin{subfigure}{0.49\textwidth}
        \includegraphics[width=\linewidth]{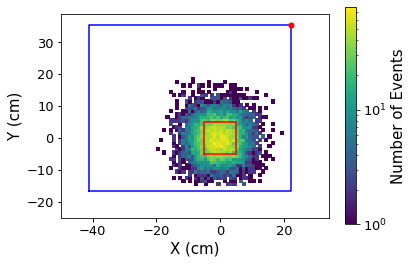}
        \caption{}
        \label{fig:run300 nHits heatmap}
    \end{subfigure}
    \hfill
    \caption{The blue rectangle is the ACD top panel's outline. The red square is the tracker outline. The red dot is the location of the top panel SiPMs. \ref{fig:run300HeatMap}: The average angle-weighted energy deposit in the ACD top panel in 1 $\times$ 1 cm bins. \ref{fig:run300 nHits heatmap}: The number of events in each 1 $\times$ 1 cm bin decreased away from the tracker, which complicated the measurement of the attenuation length. }
\end{figure}

In addition to the veto rate, this source run contained enough statistics to measure the light attenuation across the top panel. Using vetoed events with at least 4 hits in the tracker and a reduced $\chi^2$ fit to a line of at most 0.01, we measured the average energy deposit multiplied by $cos(\theta)$ in 1cm x 1cm bins across the top ACD panel using the tracks to identify the ACD interaction positions and azimuthal angle $\theta$. The factor of $cos(\theta)$ compensates for increases in energy deposit at steeper incidence angles. The average angle-weighted energy deposits were fit by an exponential decay function to provide an estimate comparable to what was found in Section \ref{sec:pre-integration}. The results are shown in Figures \ref{fig:run300LossScatter} and \ref{fig:run300HeatMap}. The estimated attenuation length from the line of best fit is 537 $\pm$ 169 cm, which is consistent with the estimate from Section \ref{sec:pre-integration} and the value quoted by Eljen for EJ 208. The main challenge with this measurement was the lack of hits away from the tracker. Figure \ref{fig:run300 nHits heatmap} shows the number of hits in each 1 $\times$ 1 cm bin used for this analysis, and it demonstrates that where there were more outliers, there was also the least statistics. What complicates the matter is that these regions were also the closest and farthest from the SiPMs, so they would have been most informative for the attenuation length.

The light collection loss across the panel was not an issue during either of the pre-flight runs. The ComPair DEE does include light collection inefficiencies for the simulations, so seeing that the measured veto rates match the simulations is evidence that the losses are not an issue for a ComPair-sized ACD with this read out scheme.

\section{Balloon Flight}\label{sec:flight}

\subsection{Overview}

\begin{figure}
    \centering
    \includegraphics[width=0.5\linewidth]{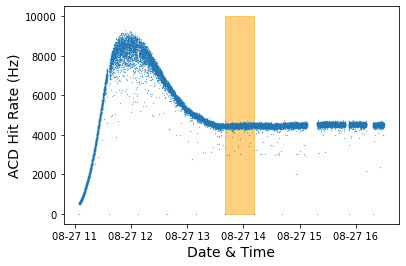}
    \caption{The event rate initially increased during ascent until the balloon passed the Pfotzer maximum, where there is maximal secondary charged particle production in the atmosphere. After this point, the event rate decreased until flattening out at a float altitude of 133,000 ft. Towards the end of the flight, the PDU's temperature exceeded its normal operating range, so the analysis will focus on the time highlighted in orange.}
    \label{fig:Flight Rates}
\end{figure}

ComPair launched on a high altitude balloon flight on August 27, 2023 from Fort Sumner, New Mexico. The following analysis used 30 minutes of data collected at approximately 133,000 ft (40.5 km) above sea level. The dataset chosen was after 2 hours of flight including ascent to consider a steady-state situation. Earlier segments contained a variable particle background due to the rapidly changing altitude, while the ComPair power distribution unit's (PDU) temperature exceeded its normal operating range later in the flight. This time slot contains significant statistics for analysis and avoids both of those complications and is highlighted in Figure \ref{fig:Flight Rates}.

\subsection{Performance}

\begin{figure}
    \centering
    \includegraphics[width=0.9\linewidth]{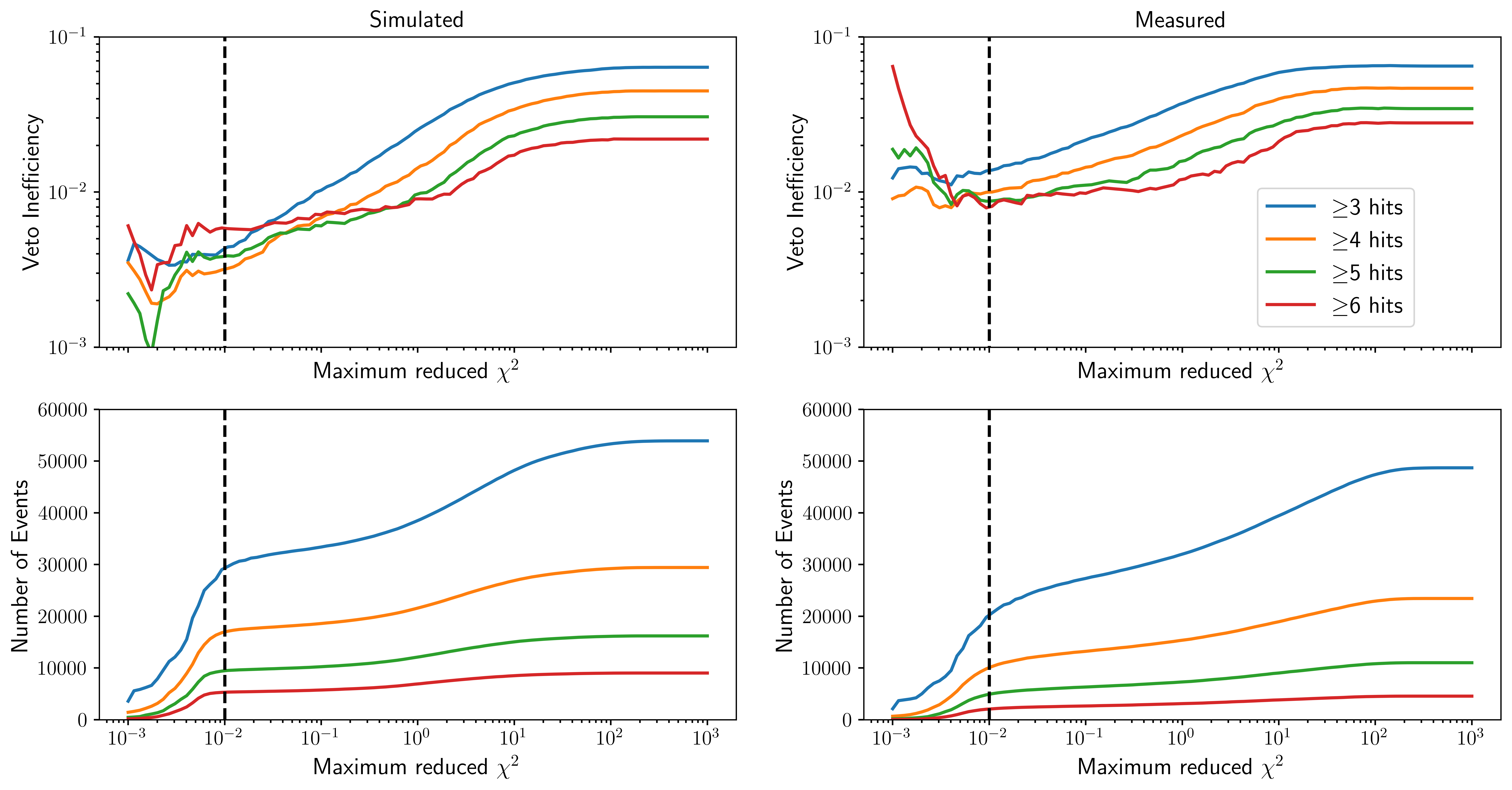}
    \caption{The veto inefficiency for both a simulated and measured ambient particle background in Fort Sumner, New Mexico in August 2023 at 133,000 ft above sea level was calculated identically to the ground-level ambient particle background and the Cs-137 source. Each solid line corresponds to a minimum number of tracker hits required per event. The vertical, dashed line represents the reduced $\chi^2$ value used to determine the ACD veto rate. The bottom row shows the number of events meeting both constraints. Both the simulated data set and the measured data set cover 30 minutes.}
    \label{fig:flightVeto}
\end{figure}

The ACD performance in flight is evaluated similarly to the pre-flight data collection runs. Figure \ref{fig:flightVeto} shows the veto inefficiency and number of eligible events for maximum reduced $\chi^2$ fit to a line and minimum number of hits in the tracker for both measured and simulated data. The veto rate decreased relative to the pre-flight tests by 0.7\%, even if the overall shape of the veto inefficiency curves remained similar.  There are several possible sources for this deficiency. One possibility is related to the difference in relative size of the noise pedestal and the charged particle spectra for the top panel compared to the side panels. In the ground-based tests, a large fraction of the charged particle flux was overhead and passed through the top panel. At float altitudes, a larger portion of charged particle flux was through the side panels, so this may point issues related to losing some charged particle interactions in the noise pedestal. Other possibilities could be in the ROSSPAD's ability to record ACD triggers at high event rates or in the data analysis pipeline's event alignment efficiency at high event rates. The ACD hit rates during pre-flight validation were between 200 and 250 Hz, while the rates during flight were about 5000 Hz. The veto rates for the pre-flight validation runs and the flight data are shown in Table \ref{Tab:Veto Rates}.

\begin{table}[ht]
\caption{We summarize the veto rates for events with reduced $\chi^2$ values $<$ 0.01. The veto rate is defined as 1 minus the veto inefficiency. The veto inefficiencies were presented in Figures \ref{fig:run286Veto}, \ref{fig:run300Veto}, and \ref{fig:flightVeto}.}
\begin{center}       
\begin{tabular}{|l|l|l|l|l|} 
\hline
 & \multicolumn{4}{c|}{\textbf{Veto Rate \%}} \\
\hline
\textbf{Run} & \multicolumn{2}{c|}{Data} & \multicolumn{2}{c|}{Simulation}\\
\hline
\textbf{Maximum Hits} & 3 & 4 & 3 & 4  \\
\hline
Ground Background & 99.3 & 99.7 & 99.4 & 99.7  \\
\hline
Cs-137 Source & 99.3 & 99.6 & 99.4 & 99.8 \\
\hline
Flight Background &	98.7 &	99.0 &	99.6 &	99.7 \\
\hline
\end{tabular}
\label{Tab:Veto Rates}
\end{center}
\end{table}

We attempted to measure the attenuation length in a similar manner to the Cs-137 run in Section \ref{sec:pre-flight}, but the results were inconclusive. The difficulties discussed above worsened at a higher event rate due to increased sampling of the noise pedestal as well as an increased likelihood of multiple particles interacting with a panel during read out.
\\ \vspace{1mm}\\
\subsection{Challenges}

There were two complications related to the balloon flight. The first was that the Arduino stopped recording data several minutes into flight. The Arduino was used for aligning interactions in the ACD with interactions in other subsystems. Without the data recorded by the Arduino, this essential step of the pipeline is not possible as originally planned. A solution using redundant data recorded by the Trigger Module was found, with the only impact on the data being UTC times are no longer directly assigned to ACD triggers. It is unclear what caused the Arduino to stop recording data, and this same issue arose for the CsI calorimeter's Arduino. The CsI calorimeter and the ACD have identical electronics. Interestingly, the CsI calorimeter's Arduino stopped recording around minute 12, whereas the ACD Arduino stopped around minute 5. We have ruled out thermal and vacuum concerns, since each Arduino was extensively tested prior to flight. The cause of the Arduino's failure is still an open question.

Another challenge that was found during post-flight analysis was a change in the spectral response for one of the SiPMs. The spectrum looked normal during ascent, but the peak moved to lower ADC values as the altitude increased and the charged particle peak disappeared altogether. At some point, the peak returned to its normal position. A possible cause is thermal contraction and expansion of the WLS or panel. During ascent, the instrument cooled down from $\sim$ 30 \textdegree C to $\sim$ 0 \textdegree C, and during this time the gain fell. This suggests that the optical coupling between either the panel and the WLS or the WLS and the SiPM decreased. Furthermore, the ComPair tracker produced a lot of heat, which eventually heated the ACD up to $\sim$ 40 \textdegree C, and over time the gain on this channel increased to near its starting value. However, this only happened to a single channel, so this effect was not widespread. Also, prior to flight we tested the change in the spectra when the SiPMs were no longer in contact with the WLS in the case that WLSs contracted but noticed no change in the signal. A possible explanation is that a gap formed between the WLS and the panel, but this is unlikely since the panel rested on top of the WLS for this channel. Despite losing one SiPM, the ACD continued to veto charged particles at a high rate, which attests to the redundant nature of the read out scheme using two SiPMs per panel.

\section{Conclusions}\label{sec:conclusion}

The ComPair ACD is a 5-panel plastic scintillator active shield to veto the charged particle background during the ComPair balloon flight. The scintillation light from each panel is collected by two WLSs that each guide the light towards a SiPM. Since the SiPMs are located near one corner of the panel, the light collection efficiency was tested using a large test panel, and the attenuation length was found to be 369 $\pm$ 39 cm, which is consistent with the value reported by Eljen. Another test of the light loss across the panel was performed during a 6.5 hour data collection run with a Cs-137 source and found the attenuation length to be 537 $\pm$ 169 cm. Two tests of the veto rate for the fully-integrated ACD were performed, one with the ambient background and another with a Cs-137 source. The tests found that the veto rate when selecting events with a reduced $\chi^2 < 0.01$ and $\ge4$ hits in the tracker were 99.7\% and 99.6\%, respectively. This is similar to the simulated veto rates in the same conditions of 99.7\% and 99.8\%, respectively.

The ComPair ACD performed well during the Fort Sumner balloon campaign. The veto efficiency was 99\% for events with $\ge4$ interactions in the tracker and a reduced $\chi^2 \le 0.01$. The flight was not without its challenges, including losing and regaining signal from one of the SiPMs, the Arduino failing to record data, and a 0.7\% reduction in veto efficiency compared to pre-flight tests. The loss of a channel is possibly explained by thermal contraction and expansion, while the Arduino's failure remains a mystery. The veto efficiency loss suggests the need to improve the ACD readout scheme at higher event rates. One possible improvement to the ACD readout scheme is to record each SiPM independently. This would reduce the size of the noise pedestal, because a SiPM will not read out if it is not triggered. Such a readout scheme may complicate event alignment, because each SiPM would have to be aligned separately.

The ComPair team is continuously upgrading the data analysis and simulations pipelines, which will provide (among other improvements) alternative measures of veto rate. For example, the method presented in this proceeding relies solely on the ComPair tracker and assumes charged particles follow straight paths and gamma rays do not. This current method ignores the energy deposited in each interaction and ignores interactions in the calorimeters. With forthcoming analysis pipeline developments, we will use Revan\cite{2006NewAR..50..629Z}, MEGAlib's Real event analyzer, to sort events into gamma-ray events and charged particle events taking energetics and the calorimeters into account.

Finally, our pre-flight testing showed that there are internal losses across the test panel, which may need to be taken into consideration for larger instruments like AMEGO. The size of an AMEGO ACD panel is 134 $\times$ 87 $\times$ 1.5 cm$^3$ with a hypotenuse of ~160 cm\cite{mcenery2019allsky}, twice as large as the diagonal of the top panel of the ComPair ACD. The integrated tests in Fort Sumner also showed limited evidence of losses across the top ACD panel. The test for losses while integrated relied on pointing charged particle tracks with the tracker onto the surface of the top panel, but there were not enough statistics in the regions farthest from the detector stack. In a future iteration of ComPair\cite{caputo}, the ACD will be able to test this better, because the detector stack will be approximately 4 times larger in area, which will increase the number of particles tracked through the regions nearest and farthest from the SiPMs. An alternative solution that would mitigate any concern for losses within the ACD panels would be to duplicate the ACD readout on the opposite corner of each panel.

\acknowledgments 

The material is based upon work supported by NASA under award number 80GSFC21M0002. This work was sponsored by NASA Astrophysics Research and Analysis (APRA) grants NNH14ZDA001N\hyp{}APRA and NNH21ZDA001N\hyp{}APRA. D. Shy is supported by U.S. Naval Research Laboratory’s Karles Fellowship.

\bibliography{report} 
\bibliographystyle{spiebib} 

\end{document}